\documentclass[aps,prl,showpacs,twocolumn,groupedaddress]{revtex4}
\usepackage{amssymb}
\usepackage{amsmath}
\usepackage{epsfig}
\usepackage{graphicx}
\graphicspath{{../tesis/eps/}{eps/}}

\newcommand{\rmd}{\mathrm{d}}
\newcommand{\rmH}{\mathrm{H}}

  \index{}
\newcommand{\Eref}[1]{Eq.~(\ref{#1})}

\providecommand{\openone}{\leavevmode\hbox{\small1\kern-3.8pt\normalsize1}}

\newcommand{\mcD}{\mathcal{D}}

\def\>{\rangle}
\def\<{\langle}

\renewcommand{\rmH}{H}
\renewcommand{\rmd}{d}

\newcommand{\la}{\langle}
\newcommand{\ra}{\rangle}
\newcommand{\lla}{\left\langle}
\newcommand{\rra}{\right\rangle}

\begin{document}

\title{Surprising relations between parametric level correlations and
 fidelity decay}

\author{H.~Kohler$^{1}$, I.~E.~Smolyarenko$^{2}$, C.~Pineda$^{3}$,
 T.~Guhr$^{1}$, F.~Leyvraz$^3$, T.H.~Seligman$^{3}$}

\affiliation{
$^1$ Fachbereich Physik, Universit\"at Duisburg--Essen, D-47057
 Duisburg, Germany\\
$^2$ School of Information Systems, Computing and Mathematics,
Brunel University, Uxbridge UB8 3PH, UK\\
 $^3$ Instituto de Ciencias F\'\i sicas, Universidad Nacional
Aut\'onoma de M\'exico,
   Cuernavaca, Mexico
}

\date{\today}

\begin{abstract}
Relations between fidelity decay, cross form--factor ({\em i.e.},
parametric level correlations), and level velocity correlations
are found both by deriving a Ward identity in a two-matrix model,
and by comparing exact results, using supersymmetry techniques, in
the framework of random matrix theory. A power law decay near
Heisenberg time, as a function of the relevant parameter, is shown
to be at the root of revivals recently discovered for fidelity
decay. For cross form--factors the revivals are illustrated by a
numerical study of a multiply kicked Ising spin chain.
\end{abstract}
\pacs{03.65.Sq, 03.65.Yz, 05.30.Ch, 05.45.Mt}
\keywords{Fidelity, Parametric Correlations, Random Matrix Theory}
\maketitle

Fidelity decay presently attracts considerable attention
\cite{refloschetal}. It measures the change of quantum dynamics of
a state under a modification of the Hamiltonian. In quantum
information, fidelity measures the deviation between a
mathematical algorithm  and its physical implementation.
From a different point of view, important insight into the properties of the underlying
systems is provided by the studies of correlations between spectra of random and/or chaotic
Hamiltonians which differ by a parameter-dependent perturbation~\cite{genpar}.
Since statistical properties of fidelity decay in random/chaotic
systems involve both spectra and eigenfunctions of the original and perturbed
Hamiltonians, existence of any connections between fidelity and
purely spectral correlations is not {\em a priori} obvious.

Random Matrix Theory (RMT) has been successful in
describing quantum many-body systems and as model for
the spectral properties of single particle systems whose classical
analogue is chaotic \cite{gmgw98}. Within RMT fidelity was
analyzed in linear response approximation \cite{1367-2630-6-1-020}
and both fidelity \cite{rudiSupersym,gorin:244105etal,sto06} and
parametric correlations \cite{sim93} were calculated exactly using
the supersymmetry method. An unexpected fidelity revival at Heisenberg time
was encountered \cite{rudiSupersym} within RMT and confirmed in a
dynamical coupled spin chain model \cite{pineda:066120}.

Earlier, differential relations between parametric spectral
correlations and parametric density correlations were established
\cite{taniguchi1,taniguchi2}. By relating the
latter to the fidelity amplitude {\em via} Fourier transform, we
show in this letter that the existence of these relations opens a
crucial insight into the properties of fidelity decay. By
analyzing the characteristic features of the parametric
correlations in the time domain, the cross--form factor, we
discover a new, simple interpretation of the previously puzzling
phenomenon of revival~\cite{rudiSupersym}.  These
relations follow directly from the basic definitions and
symmetries of the underlying matrix models, being
essentially Ward identities. We show that they are valid under very 
general assumptions. No explicit (e.g., supersymmetric)
calculation is required, however they rely on the universality of
the parametric spectral correlations at the scale of mean level
spacing. We thus explain the origin of various relations
connecting spectral and wave-function correlations, and
establish a unified framework for their analysis and
generalizations. A relation between fidelity decay and level
velocity correlation function is given. The latter is important from the
experimental point of view, being used for independent access to
system parameters. We confirm the general results comparing
fidelity decay and cross--form factors in RMT. We  illustrate our 
analytical results with a numerical
study of a multiply kicked Ising spin chain.

We consider Hamiltonians modeled by $N\times N$ matrices
\begin{equation} \label{eq:Hamiltonian}
H^{\pm}(\lambda)\ =\ H \pm \lambda V/2 \ ,
\end{equation}
where $H$ and $V$ are independently drawn from ensembles of the
same symmetry. In particular, $V$ is drawn from the GOE, the GUE
or the GSE ensembles of RMT, labeled $\beta=1,2,4$. The ensemble
average over both is indicated by angular brackets.  It
is convenient to fix the variances as
$\la H_{ij}H_{kl} \ra $ $= D^{-1} \la V_{ij}V_{kl}\ra$
where $D$ is the mean level spacing of $H^{\pm}(0)$ in the
 energy region of interest.
In the RMT case, $D =\pi^2/N$ in the center of the spectrum. The
mean level spacing is then $\lambda$-independent up to corrections
of order $1/N$. By construction, $H^{\pm}(\lambda)$ is in the same
symmetry class as $H$ for any $\lambda$.

The parametric two--level correlation function is defined as
\begin{equation}\label{pcdef}
\widetilde{R}_\beta(E^{+},E^{-},\lambda) = \sum_{n,m}\la
\delta(E^{-}-\epsilon^-_n(\lambda))
 \delta(E^{+} -\epsilon^+_m(\lambda))\ra \ .
\end{equation}
It is mapped onto a dimensionless energy scale, where the mean
level spacing is rescaled to unity. One has
\begin{equation}
\label{loccorr} \widetilde{X}_\beta(r,\lambda) = \lim_{N\to
\infty} D^2 \widetilde{R}_\beta(E^{+},E^{-},\lambda) \ ,
\end{equation}
which solely depends on the difference $r= (E^{+}-E^{-})/D$. The cross 
form-factor is obtained as a Fourier transform
\begin{equation}
\widetilde{K}_{\beta}(t,\lambda) =
\int_{-\infty}^{+\infty}
           \left(
1-\widetilde{X}_\beta(r,\lambda)\right)
 e^{2\pi \imath t r} \rmd r\, ,\ t>0 \ .
\label{ftcorr}
\end{equation}
Time $t$ is measured in units of Heisenberg time $t_\rmH= D^{-1}$.
Fidelity decay is expressed {\em via} the echo
operator  \cite{refloschetal}
\begin{equation}
M(t,\lambda)\ = \exp(\imath 2\pi t H^{-}(\lambda)/D)\,
\exp(-\imath 2\pi t H^{+}(\lambda)/D) \ .
\end{equation}
Its expectation value with a given state is the fidelity amplitude and
its average
\begin{equation}
\label{fidgen}
\quad f_\beta (t,\lambda)  =  \frac{1}{N} {\rm tr}\langle
M(t,\lambda)\rangle
\end{equation}
is a measure for the difference in the two time evolutions as
a function of $\lambda$.

\begin{figure}
\begin{center}
\includegraphics{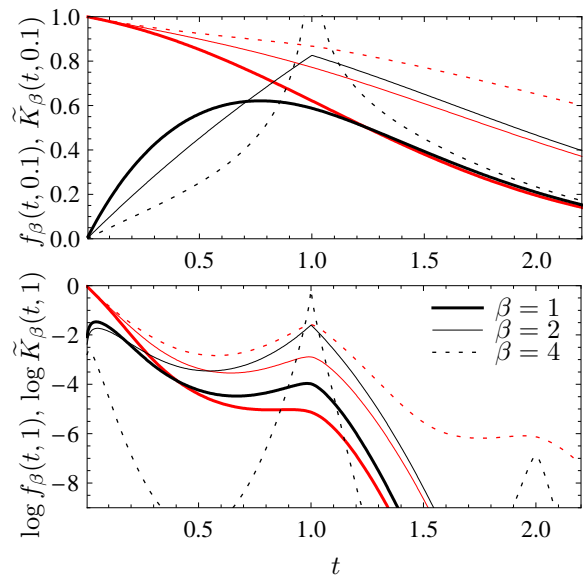}
\end{center}
\caption{\label{fig1} (Color online)  Cross form-factor (black) and
  fidelity (red) versus time for two different values $\lambda$.  The
  results for the three Gaussian ensembles $\beta=1,2,4$ are given as
  thick solid, thin solid and dashed lines, respectively.}
\end{figure}

The functions in Eq.~(\ref{loccorr}) were calculated exactly with
the supersymmetry method \cite{sim93} for $\beta=1,2,4$. The
Fourier transforms are (see also \cite{eduardo})
\begin{multline}\label{gueresult}
\widetilde{K}_{1}(t,\lambda) = \int_{\max(0,t-1)}^t
\rmd u \int_0^u \rmd v
 \frac{2t^2(t-u)(1-t+u)}{(v^2-t^2)^2} \\
\shoveright{
\times
\frac{\exp\left(-2\pi^2\lambda^2\left[2ut+t-t^2+v^2\right]\right)}
{\sqrt{(u^2-v^2)(u^2+2u+1-v^2)}},}   \\
\widetilde{K}_{2}(t,\lambda) =
\frac{\exp\left(-2\pi^2\lambda^2t^{1+\theta(t-1)}\right)}{2\pi^2
 \lambda^2 t}
       \sinh\left(2\pi^2\lambda^2 t^{2-\theta(t-1)}\right)\ , \\
\shoveleft{
\widetilde{K}_{4}(t,\lambda)=
 t^2\int_{-1}^{+1} \rmd u \int_0^{1-|u|} \rmd v
\frac{(u+t)^2-1}
  {(t^2-v^2)^2}} \\
 \times  \frac{v\theta(u-1+t)
\exp\left(-\pi^2\lambda^2 [ t^2-v^2+2tu] \right)}
{\sqrt{\left[(u-1)^2-v^2\right]\left[(u+1)^2-v^2\right]}}
,
\end{multline}
with Heaviside's $\theta$--function. 
For $\lambda=0$ the cross form-factors reduce to the standard form
factors $K_\beta(t)$ \cite{gmgw98}, i.e.~$\widetilde{K}_{\beta}(t,0) =
K_\beta(t)$. In Fig.~\ref{fig1} we show $\widetilde{K}_{\beta}(t,\lambda)$ versus time $t$ for two values of
$\lambda$.  For $\lambda=0.1$, the correlations vanish as
$t\to\infty$. A second peak develops in the GSE case for
$\lambda=1$ at $t=2$.  The singularity at $t=1$
persists.  For the GOE and for the GUE cases finite peaks appear at
$t=1$ but not at multiples thereof. For all ensembles another peak
appears for small times $t\ll 1$.  Its location scales asymptotically
with $\lambda^{-2}$.

The peak appearing at $t=1$ for large $\lambda$ clearly
indicates that here the correlations decay more slowly as a
function of $\lambda$ than at all other times $t$.  We study this in
more detail by an asymptotic analysis in $\lambda$ of the exact
integral expressions (\ref{gueresult}). We calculate the weight $
W_\beta(t,\lambda)$ $=$
$\lambda^2\int_{t-\lambda^{-2}}^{t+\lambda^{-2}}
\widetilde{K}_{\beta}(t^\prime,\lambda) dt^\prime$ of the peaks at
$t=1$ and, for the GSE, also at $t=2$.  In contrast to the peak height
the weight is well defined for all times for all three
ensembles.  We find 
\begin{equation}
\label{asym1}
 W_\beta(1,\lambda) \ \propto \ \lambda^{-2(4-\beta)/\beta} + {\cal
O}(\lambda^{-8/\beta})
\end{equation}
and $W_4(2,\lambda)\propto\lambda^{-6}$.  The weight of the first peak
$t\approx 0$ scales as $\lambda^{-2}$ independently of the ensemble.
These decays are governed by power laws in $\lambda$ while they are
exponential for all other times. We shall see below, that the behavior of the 
cross form-factor at $t=1$ is directly related to fidelity revivals, which
for the GSE also occur at $t=2$.

For the classical ensembles $K_{\beta}(t)$, is non--analytic at $t=1$ 
\cite{gmgw98}. The degree $\mcD(g,x)$ of
non--analyticity of a function $g(x)$ at $x$ is defined as the smallest
integer $\mcD$ for which the $\mcD$-th derivative $g^{(\mcD)}$ is
discontinuous at $x$. For the form factor we find $\mcD(K_{4},1)= 0$,
$\mcD(K_{2},1)=1$ and $\mcD(K_{1},1) = \mcD(K_{4},2) = 3$.  For typical
times we find $\mcD(K_{\beta},t) = \infty$, because $K_{\beta}(t)$ is analytic.
We thus arrive at a relation between the asymptotic
behavior of $W_\beta(t,\lambda)$ for large perturbation to the degree
of non--analyticity of $K_{\beta}(t)$ which reads
\begin{equation}
\label{asym2}
 W_\beta(t,\lambda) \ \propto \
\lambda^{-2\mcD(K_{\beta},t)} + \ldots \, , \ t>0 \
.
\end{equation}
We conjecture that this relation also holds for arbitrary
$\beta\neq 1,2,4$.

We use the multiply kicked Ising (MKI) spin chain proposed in
\cite{prosenKI, pineda:066120} to illustrate the revival in the cross
form-factor. The MKI spin chain is a periodic 1-d array of $L$ spins
$1/2$ with anti--ferromagnetic nearest--neighbor Ising interaction of
unit strength and periodic boundary conditions.  Each spin receives
periodically two different kicks of instantaneous magnetic field
pulses. The time--reversal breaking Floquet operator of the system is
$U_\text{MKI}$ $=$ $U_\text{I}U_\text{K}^{(1)}
U_\text{I}U_\text{K}^{(2)}$, where $U_\text{I}$ is the time evolution
operator of the unkicked spin chain and $U_\text{K}^{(n)}=\exp (-\imath
\sum_j \vec{b}^{(n)} \cdot\vec{\sigma}_j)$, ($n=1,2$) describes each
magnetic pulse with a dimensionless magnetic field $\vec{b}^{(n)}$.
$\vec{\sigma}_j$ are the Pauli operators for particle $j$.  The
translational symmetry ($\vec{\sigma}_j \to \vec{\sigma}_{j+1}$) which
foliates the space in $L$ different symmetry sectors. For the choice
$\vec{b}^{(1)}=(0,1,1)$ and $\vec{b}^{(2)}=(1.4,0,1.4)$ the spectral
statistics in most symmetry sectors display excellent agreement with
the GUE. We introduce an additional magnetic pulse of strength $\delta$
in $z$ direction as a perturbation. We define $U_\delta=$ $U_\text{MKI}
\cdot \exp\left( \imath\delta \sum_{j=0}^{L-1} \sigma ^z_j \right)$ and
calculate the cross form-factor of $U_\text{MKI}$ and $U_\delta$ using
direct diagonalization, omitting the problematic sectors.  The
perturbation strength $\lambda$ can be calculated from $\delta$ using
the  correlation functions of the perturbing
operator~\cite{refloschetal}. Details are given elsewhere.

\begin{figure}
\begin{center} \includegraphics{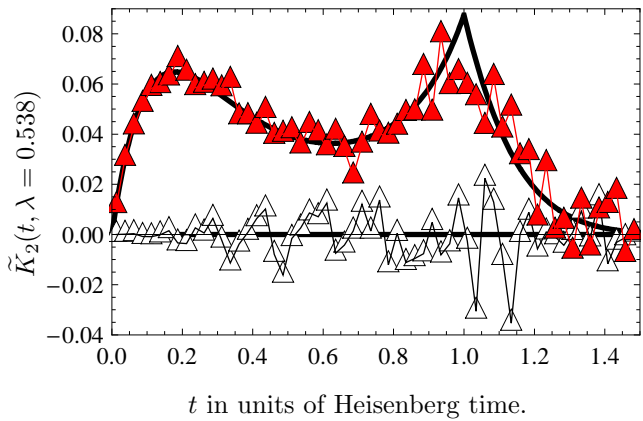}
\end{center}
\caption{(Color online) The cross correlation function for the MKI
  model with $L=18$. Filled/empty triangles correspond to the
  real/imaginary part of the cross correlation.  The statistical error
  (measured by the imaginary part) is small enough to observe clearly
  the peak.  The theoretical expectation \Eref{gueresult} is plotted as
  a thick curve. }
\label{fig:KI}
\end{figure}
In Fig.~\ref{fig:KI} we compare results of this model with RMT results
of Eq.~(\ref{gueresult}).  We see good agreement with the theoretical
result, up to statistical fluctuations, measured by the imaginary part.
In particular the peak at $t=1$ is observed.

We now derive the announced differential relations connecting
the cross form--factor with fidelity, deferring the reader to
a follow--up paper for more details. Consider a general
4-point parametric correlation function
$F_{\alpha\beta;\gamma\delta}(z_1,z_2)=\lla
\mathcal{F}_{\alpha\beta;\gamma\delta}(z_1,z_2;H^{-},H^{+})\rra$
defined as
\begin{equation}
\label{fourpoint}F_{\alpha\beta;\gamma\delta}(z_1,z_2)=
\left\langle\left(\frac{1}{z_1-H^{-}}\right)_{\alpha\beta}
\left(\frac{1}{z_2-H^{+}}\right)_{\gamma\delta} \right\rangle\ .
\end{equation}
The definition of the angular brackets is now expanded to denote either
the average over an arbitrary matrix ensemble with
measure $d\nu(H)$, or the energy averaging over a spectral window
of an individual quantum chaotic system. We do not require
$d\nu(H)$ to be Gaussian or even rotationally invariant. The
distribution of $V$, on the other hand, is required to be Gaussian in
order to ensure the existence of the announced differential
relations at finite order (see Eq. (\ref{diff}) below).
The Fourier transform of fidelity amplitude corresponds to
$F_{\alpha\beta;\beta\alpha}$ and parametric spectral correlator
corresponds to $F_{\alpha\alpha;\beta\beta}$, with a summation over
double indices. Introducing $H_{1,2}$ {\em via} $\delta(H_{1,2}-H^{\pm}(\lambda))$,
Fourier transforming the matrix delta-functions and integrating over $V$,
the averages are rewritten as
\begin{eqnarray}
\label{action}
&&\lla\mathcal{F}\rra=\int d\Lambda_1 d\Lambda_2 d\nu(H) dH_1dH_2 \nonumber\\
&& \quad e^{\mathrm{tr}\left[\imath\Lambda_1
(H_1-H)+\imath\Lambda_2(H_2-H)-\frac{\lambda^2 D^2}{4\beta}
(\Lambda_2-\Lambda_1)^2\right]}\,\mathcal{F}
\end{eqnarray}
where the symmetry class of the matrices $\Lambda_1$, $\Lambda_2$
corresponds to the symmetry class of $H$, and multiple factors of
$2\pi$ are absorbed into the definition of $d\Lambda_{1,2}$ .

The invariance of the flat integration measures $dH_{1,2}$
with respect to independent shifts in $H_1$ and $H_2$ implies
\begin{equation}
\label{Schwinger2} \lla \mathrm{tr}\left(\frac{\partial}{\partial
H_1}-\frac{\partial}{\partial H_2}\right)^2\mathcal{F}_{\alpha\alpha;\beta\beta}\rra =
-\lla\mathrm{tr}(\Lambda_1-\Lambda_2)^2\mathcal{F}_{\alpha \alpha;\beta\beta}\rra
\end{equation}

The full measure in Eq.~(\ref{action}) is also {\em approximately}
invariant under a simultaneous shift of $H_1$ and $H_2$. The
violation of this symmetry stems from the non-invariance of $d\nu(H)$
under the shifts of $H$. However, universality implies that the correlation
functions depend on such shifts only through the average density of states
and level velocity variance~\cite{sim93}. This dependence is thus manifested
only on time scales much shorter than $t_H$, which is of interest here.
In invariant unitary RMT ensembles universality under shifts was shown
in \cite{sm03}. Although not yet proved in general, no violations of this
universality are known. In particular universality follows automatically
in models which allow for field theoretical representations of correlation
functions \cite{efetov}. With these {\em caveats} we can set
\begin{equation}
\label{Schwinger3} \lla \mathrm{tr}\left(\frac{\partial}{\partial
H_1}+\frac{\partial}{\partial H_2}\right)^2
\mathcal{F}_{\alpha\alpha;\beta\beta}\rra \approx 0
\end{equation}
and combine with Eq.~(\ref{Schwinger2}) to
\begin{equation}
\label{eq:Ward1} 4\lla\mathrm{tr}\frac{\partial}{\partial
H_1}\frac{\partial}{\partial
H_2}\mathcal{F}_{\alpha\alpha;\beta\beta}\rra=\lla\mathrm{tr}
\left(\Lambda_1-\Lambda_2\right)^2\mathcal{F}_{\alpha\alpha;\beta\beta}\rra
 \ .
\end{equation}
Using
\begin{eqnarray}
\label{derivatives}
&& \lla\mathrm{tr}\frac{\partial}{\partial H_1}
\frac{\partial}{\partial H_2}
\mathcal{F}_{\alpha\alpha;\beta\beta}\rra=\frac{\partial^2}{\partial
z_1 \partial z_2} F_{\alpha\beta;\beta\alpha},\\
\label{diff}
&&\lla\mathrm{tr}(\Lambda_1-\Lambda_2)^2\mathcal{F}_{\alpha\alpha;\beta\beta}\rra
=-\frac{4\beta}{D^2} \frac{\partial}{\partial
\lambda^2}F_{\alpha\alpha;\beta\beta}
\end{eqnarray}
and Fourier transforming $F$, we finally show that
\begin{equation}
\label{identity1} \frac{\partial}{\partial \lambda^2}
\widetilde{K}_\beta(t,\lambda)= -\frac{4\pi^2 t^2}{\beta}  f_\beta(t,\lambda),
\end{equation}
under very general assumptions. The averaged fidelity amplitude 
has been calculated in Ref.~\cite{rudiSupersym} for the GOE and the GUE. For the GSE,
\begin{multline}
\label{fidgse}
 f_4(t,\lambda)= \int_{-1}^{+1}\rmd u
\int_0^{1-|u|} \rmd v
 \frac{(u+t)^2-1}{(t^2-v^2)^2}|v|\theta(u-1+t) \\
\times  \frac{(t^2-v^2+2tu)
\exp\left(-\pi^2\lambda^2\left[t^2-v^2+2tu\right]\right)}
{\sqrt{\left[(u-1)^2-v^2\right]\left[(u+1)^2-v^2\right]}}
\ .
\end{multline}
A direct comparison of the exact expressions for
$\widetilde{K}_\beta(t,\lambda)$ obtained in the present
contribution and of the ones for $f_{\beta}(t,\lambda)$ in
Ref.~\cite{rudiSupersym} and in Eq.~(\ref{fidgse}) confirms the
validity of Eq.~(\ref{identity1}) in the universal RMT regime
(although we stress that it is valid for any
disordered/chaotic model which exhibits a separation of scales
between the oscillatory local and smooth global behavior of
spectral statistics).

Relation (\ref{identity1}) allows us to view fidelity revival at
Heisenberg time $t_H$ as being rooted in the algebraic decay of the
cross form-factor.  Furthermore, due to the established relations,
power law decay as a function of $\lambda$ must also hold for
fidelity at $t_H$ and, for the GSE, at $2t_H$. 
This could also have been derived directly from
the exact equations.

In Fig.~\ref{fig1}, we show the fidelity amplitude.  Similar to the
behavior of the cross form-factor, a peak at $t=1$
appears for all three ensembles \cite{rudiSupersym} and for increasing
$\beta$ the peaks become more and more pronounced. In the GSE case, a
second peak emerges at $t=2$. This peak was not
seen in the numerics of \cite{rudiSupersym} as it was beyond numerical
accuracy.

Relation (\ref{identity1}) is, essentially, a Ward identity
associated with the action (\ref{action}). It immediately allows
to establish a connection between fidelity amplitude and the Fourier transform 
of the level velocity correlator $C(t,\lambda)$, which 
is related to $\widetilde{K}$ by a Ward identity $ -4\pi^2
t^2C(t,\lambda)=(\partial^2/\partial \lambda^2)
\widetilde{K}(t,\lambda)$ (see, {\em e.g.}, \cite{lerner}).
As seen from Eq.~(\ref{action}), $\widetilde{K}$ is a function of $\lambda^2$,
it follows from  (\ref{identity1}) after a short calculation that
\begin{equation}
\beta
C(t,\lambda)=\left(2+4\lambda^2\frac{\partial}{\partial
\lambda^2}\right)f(t,\lambda) \ .
\end{equation}

To summarize, we established relations between cross form-factor
and level velocities on the one hand, and fidelity decay on the
other hand. They hold in any system displaying universality of
spectral correlations. The present formalism can be used to
construct a whole family of Ward identities relating apparently
unconnected correlation functions. One instance are
generalizations of the `optical theorem' found in
\cite{taniguchi2}, which relates fidelity amplitude for small
perturbations to the spectral form factor $K_\beta(t)$.
Further, the results presented here do not apply to crossover
regimes, where $V$ changes the symmetry of $H$. One such relation
was obtained using supersymmetry methods in \cite{taniguchi1}. A
broader set of differential relations, generalizing those of
\cite{taniguchi1}, can be obtained by utilizing different
transformation properties of the action Eq.~(\ref{action}) under
symmetry-preserving and symmetry-violating shifts. Details of
these and other hierarchies of relations will be presented
elsewhere.

Our findings make it possible to explain features of one
quantity via the other, i.e.~the characteristics of fidelity decay in
terms of the cross form-factor or vice versa.  In particular, the
revivals of both quantities are linked in this way.  We studied in
detail the decay laws of the corresponding peaks.  Further peaks are
not possible.  The very occurrence of the peaks in the cross
form-factors is neither trivial nor intuitive and will be discussed in
elsewhere.

\begin{acknowledgments}
We thank T. Gorin, J. Keating, T. Prosen, D. Savin, R.~Sch{\"a}fer and
H.~J. St{\"o}ckmann for useful discussions. We acknowledge grants
DFG--KO3538/1-1 (HK), DFG--SFB Transregio 12 (TG, HK), EPSRC--EP/E037429/1 (IS), 
PAPIIT IN112507 and CONACyT 57334 (TS, CP).
\end{acknowledgments}

\end{document}